\documentclass[doublecol,amsmath,amssymb,cite]{epl2}
\usepackage{graphicx}% Include figure files
\usepackage{dcolumn}% Align table columns on decimal point
\usepackage{bm}% bold math
\newcommand{\gammadot}{\dot{\gamma}}

\title{Particle Diffusion in Slow Granular Bulk Flows}

\shorttitle{} %Insert here a short version of the title if it exceeds 70 characters

\author{Elie Wandersman\inst{1,2}, Joshua A. Dijksman\inst{1,3} \and
Martin van Hecke\inst{1}}

\shortauthor{E. Wandersman \etal}

\institute{
\inst{1} Kamerlingh Onnes Laboratorium, Universiteit
Leiden,   P.O. Box 9504, 2300 RA Leiden, The Netherlands. \\
\inst{2} Laboratoire Jean Perrin,  FRE 3231, Universit\'{e} Pierre et Marie Curie - CNRS, Ecole Normale Sup\'erieure, 24 rue Lhomond, 75005 Paris, France\\
\inst{3} Department of Physics, Duke University, Science Drive,
Durham NC 27708-0305, USA\\}

\pacs{83.80.Fg}{Rheology of Granular materials }
\pacs{47.57.Gc}{Granular flows} \pacs{83.80.Ab}{Rheology of
Glasses}

\abstract{ We probe the diffusive motion of particles in slowly
sheared three dimensional granular suspensions. For sufficiently
large strains, the particle dynamics exhibits diffusive Gaussian
statistics, with the diffusivity proportional to the local strain
rate --- consistent with a local, quasi static picture.
Surprisingly, the diffusivity is also inversely proportional to
the depth of the particles within the flow
--- at the free surface, diffusivity is thus ill defined. We find
that the crossover to Gaussian displacement statistics is governed
by the same depth dependence, evidencing a nontrivial strain scale
in three dimensional granular flows.}

\begin{document}

\maketitle
\date{\today}

%\pacs{45.70.Cc,81.05.Rm,45.70.-n} \keywords{granular flow,
%suspensions, packing fraction, inertial number, index matching}
%\maketitle

Flowing soft disordered solids, such as colloidal glasses and gels
\cite{WeitzPRL10, SchallScience07,NIPADurianPRL2010}, foams and
emulsions \cite{GoyonNature08,KatgertPRL08,TighePRL10} or granular
media \cite{ForterreAnnRevFlu08, GDRMidi,
LosertPRL00,PouliquenPRL04,BehringerPRE04,FenisteinPRL06,DijksmanPRE10},
exhibit rich particle dynamics and complex rheology. In the last
decade, major progress has been made in the understanding of the
rheology of granular media and suspensions. Important developments
include the description of rapid, ''inertial'' dry granular flows
\cite{ForterreAnnRevFlu08,GDRMidi}, and the observation of
connections between this inertial rheology and the classical
rheology of density matched suspensions \cite{BoyerPRL11}. The
most challenging regime is that of very slow flows, where the
stresses become (approximately) rate independent
\cite{DijksmanPRL11} and nonlocal effects play an important role
\cite{NicholPRL10,PouliquenPRL10,KamrinPRL12}. Little is known
about the fluctuations of the microscopic grain motion, i.e.,
self-diffusion, in such flows. A recurring problem is that
granular media are opaque, so only motion in two dimensional (2D)
model systems \cite{BehringerPRE04}, near transparant walls
\cite{LosertPRL00,clement08} or at a free surface can be observed
\cite{PouliquenPRL04}. Walls lead to layering
\cite{MuethNature2000} and other artifacts \cite{SchallARFM10},
while at the free surface, the particles experience a different
local geometry than in the bulk. Measurements in the bulk of a
granular flow are thus essential.

Recently, we have shown \cite{DijksmanPRE10} that for sufficiently
slow flows, dry granular media and non-density, but optically
matched suspensions exhibit identical  flow patterns and rheology,
opening up the possibility to probe 3D microscopic particle
dynamics within slow granular flows. Here we experimentally probe
the full 3D particle trajectories of a sheared granular suspension
in a split bottom cell (Fig.~1). We focuss on sufficiently slow
strain rates so that viscous effects are negligible: as detailed
below, typical Reynolds and Stokes numbers are of order $10^{-4}$
and $10$, respectively, implying that we are in the so-called free
fall regime, relevant for dry granular flows
\cite{DijksmanPRE10,CourrechPRL03}.

In a quasistatic picture, the mean squared displacements should
only depend on the strain, or equivalently, the diffusion
coefficient should be proportional to the strain rate, as seen
already in \cite{BehringerPRE04}. By extracting the particle
diffusivity and local strain rates from our particle trajectories,
we confirm that this is true in 3D also.

\begin{figure}[tbhp]
\begin{center}
\includegraphics[width=0.47\textwidth]{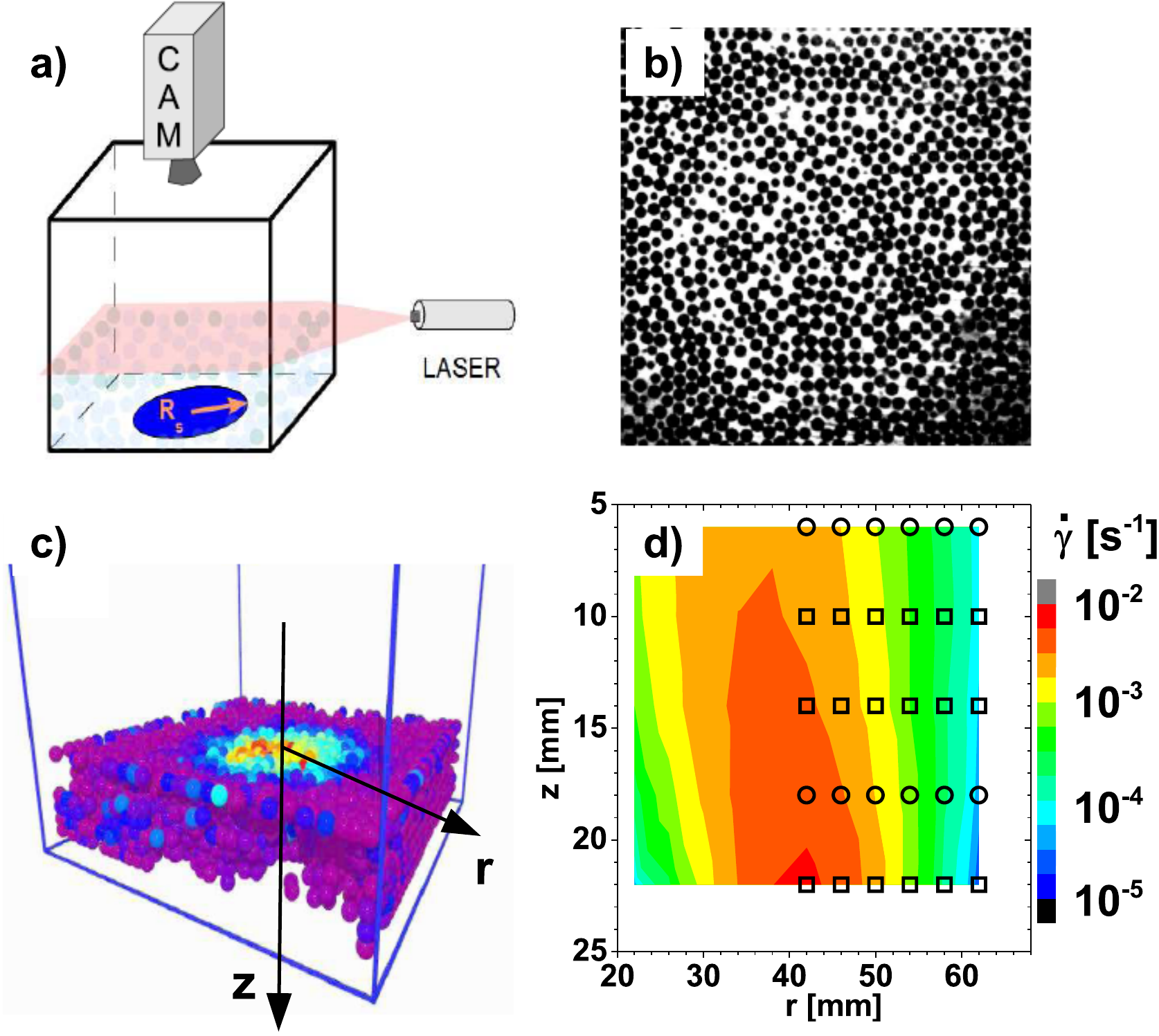}
%if the picture is bigger, gets separate page? Don't understand why.
\end{center}
\caption{\label{fig:epsart} (Color online) (a) Setup: a slowly
rotating disk at the bottom of a cubic container drives a slow
granular flow. In this experiment, we probe driving rates $\Omega$
from $0.005$ to $0.05$ rpm. The container is filled with
non-density matched suspension consisting of PMMA beads, index
matched fluid and fluorescent dye. The motion of the particles in
this granular suspension are captured in 3D by vertical scanning
of a lasersheet. (b) Example of image obtained by illuminating a
single slice and capturing the image with a CCD camera. (c) After
particle tracking, the motion of the particles can be resolved in
3D --- in this rendered image, the color of the particles
represents their instantaneous velocity. (d) Strain rate field
(for the case $\Omega=0.05$ rpm), where the color code denotes the
strain rate $\gammadot$. Symbols indicate the region in which we
will probe local strain rates and fluctuations (circles: data used
in Fig.~2abc; squares: additional data used in Fig.~2d). }
\end{figure}

The crucial point is that our 3D geometry allows us to probe the
depth dependence of the diffusivity. In the simplest picture, one
might expect the diffusion to the independent of depth. However,
our main result is showing that the diffusion is inverse
proportional to the depth, and thus diverges near the free
surface. To answer whether this depth dependence stems from a
depth dependence of the mean squared displacements or of the
characteristic strain, both of which can lead to a depth dependent
diffusivity, we probe the displacement distributions. These evolve
from very wide distributions at short time with (reduced) kurtosis
exceeding 40, to Gaussian distributions at late times. We find
that, for a given depth, the kurtosis scales as a powerlaw with
strain, and that we can collapse kurtosis data taken at different
depths by introducing a depth dependent characteristic strain
scale
--- the same depth dependent characteristic strain scale
governs the diffusivity. Taken together, the granular fluctuations
provide strong evidence for a non-trivial, depth dependent strain
scale which governs slow granular flows.

{\em Setup ---} We probe the particle trajectories in the bulk of
a gravitational (non-density matched) granular suspension sheared
in the split-bottom geometry \cite{FenisteinPRL06,DijksmanPRE10,
DijksmanSM10}. This geometry produces smooth and reproducible flow
profiles, is well studied and allows to investigate a wide range
of shear rates within a single experiment. This allows us to probe
the relation between particle displacements, $\dot{\gamma}$, and
distance to the free surface.

We use an identical geometry as the one used in
\cite{DijksmanPRE10} (see Fig. 1a). A cubic container (width 150
mm), at the bottom of which a disk of radius $R_s$ = 45 mm rotates
at an angular velocity $\Omega$ ranging from $0.005$ to $0.05$
rpm, driven by a microstepping motor.

Imaging of the suspension is performed using an index-matching
technique, as introduced in
\cite{TsaiPRL03,LosertPRL08,KingaSoftMatt10, DijksmanPRE10,
DijksmanRSI12}. The granular suspension consists of monodisperse
PMMA spheres (Engineering Labs, diameter 4.6 mm $\pm 0.1$ mm
placed in a mixture of Triton X-100, fluorescent dye (Nile Blue)
and a few droplets of a 37\% HCl solution to tune the absorption
spectrum of the dye. The fluid is prepared to closely match the
refractive index of particles. The difference in density between
the fluid and the particles is $\Delta \rho$=110 $\pm$ 5 kg/m$^3$.
The fluids viscosity is 0.30 $\pm$ 0.05 Pa.s. The split-bottom
cell is filled with grains up to a fixed height $H=25 \pm$ 2 mm;
there is always a layer of fluid above the grains, so that the
fluids surface tension does not play a role for the particle
dynamics. The suspension is illuminated with a laser sheet
($\lambda$=635 nm, power 30 mW) parallel to the bottom of the box
(Fig. 1a) and mounted on a z-stage which allows the illumination
of slices of the suspension at different heights (every 500
$\mu$m, starting 3 mm from the bottom of the box). The thickness
of the laser sheet is $\approx 200$ $\mu$m.

Image acquisition is done with a triggered 8 bit CCD camera
operating at 10 Hz. Both the camera and laser sheet are mounted on
vertical stages and can be translated synchronously (velocity
$\sim$ 5 mm /sec) so that the probed slice of the suspension
always stays in focus. Since the fluid is fluorescent but the
particles are not, a single slice shows a collection of black
disks (cross sections) on a white background (see Fig.~1b). The
contrast is sufficient to image the entire box, but best on the
half closest to the laser, where we focus our analysis. We
determine the particle trajectories using 3D tracking techniques
developed for confocal images of colloidal suspensions
~\cite{WeeksTrack96}. We follow the particles up to 1000
consecutive scans. We estimate the tracking error, obtained by
tracking particles in a stationary packing, to be 1/30 of the
diameter in the plane of the sheet and 1/10 in the vertical
direction.

\begin{figure*}[!t]
\includegraphics[width=18cm]{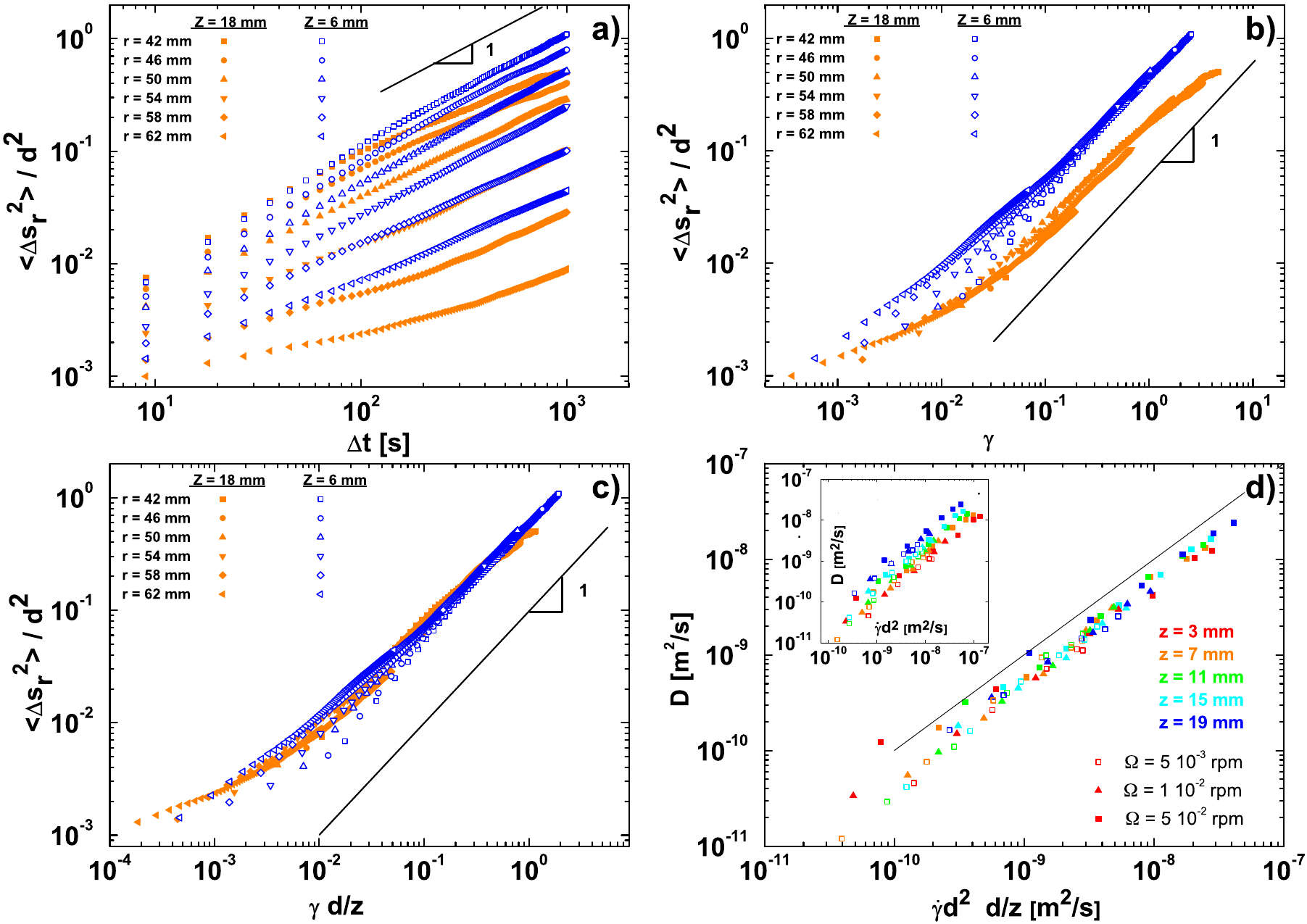}
\caption{\label{fig2} (a) Examples of non dimensional mean squared
displacements $\langle \Delta s_r^2 \rangle/d^2$ as function of
time interval $\Delta t$. $\Omega$ is fixed at 0.05 rpm, and we
only plot data at $z=7$ mm and $z=19$ mm, for several values of
$r$ as indicated (see Fig.~1d) --- for these values, the local
strain rate $\gammadot$ varies over 2.5 decades. At late times,
the curves all show a simple linear growth including diffusive
behavior, with a diffusivity which clearly grows with the local
strain rate. (b) When plotted as function of the local strain
$\gamma:=\gammadot \Delta t$, mean squared displacements for the
same value of $z$ collapse, indicating that the dynamics are
quasi-static. Families of collapsed curves are surprisingly
distinct for different $z$. At late times, the curves all show a
simple linear growth with strain, allowing the unambiguous
definition of a dimensionless diffusion coefficient as $D:=
\langle \Delta s_r^2 \rangle/d^2/\gamma$. (c) By rescaling the
local strain with the inverse of the dimensionless layer number
$z/d$, the mean squared displacements for $z=7$ mm and $z=19$ mm
can be scaled on top of each other. (d) Similarly, the values of
$D$ for all values of $z$ and $r$ (see Fig.~1d) vary linearly with
the inverse of the depth. The solid line is the prediction from
Eq.~4. Inset: $D$ as a function of $\gammadot d^2$,
from which the pressure effect on diffusion is explicit.
%{\bf
% NOte, this has to be color fig 800 euro, 150
%orso for each addl color page}
}
\end{figure*}

{\em Flow profiles and local strain rates---} From the 3D particle
tracking, we obtain a time sequence (maximum 1000 steps) of
particles positions $\vec{r}_{i}(t)$, where $i$ denotes the label
of the approximately 5000 particle trajectories which can be
reconstructed. Since the shear-rate is not homogeneous in the
split-bottom geometry, we divide the measurement area into bins.
Given the azimuthal symmetry of the problem, these bins are chosen
as squared tori (axis of revolution the vertical axis $z$) of size
$d_R=d_Z$=4 mm ($\approx 0.85~ d$). We determine the average
velocity $\left< \vec{v}(r,z)\right>$ within each bin. The flow is
orthoradial, as the transverse components $v_r$ and $v_z$ are
negligible compared to $v_{\theta}=r \omega(r,z)$.

The average angular velocity field $\omega(r,z)$ is independent of
$\Omega$ in the range used here \cite{DijksmanPRE10}. As expected
for this value of the ratio $H/R_s$ \cite{FenisteinPRL06,
DijksmanPRE10,DijksmanSM10}, the flow has a ``trumpet like''
shape, with shear bands arising from the edge of the rotating disc
and propagating within the bulk of the suspension. We successfully
fit $\omega(r,Z_0) / \Omega$ for different $Z_0$ using error
functions \cite{FenisteinPRL06, DijksmanPRE10,DijksmanSM10} (not
shown). We use these fits to get an accurate estimate for the
local strain rate within each bin, assuming that the velocity
gradient tensor has an unique non-zero component
\begin{equation}
\dot{\gamma}_{r\theta}(r,Z_0)=r/2 \frac{\partial
\omega(r,Z_0)}{\partial r}~.
\end{equation}
The resulting local strain rate is plotted in Fig.~1d. By varying
the disk rotation speed from $\Omega$=0.005 rpm to 0.05 rpm, we
access local strain rates from 10$^{-5}$ up to  $10^{-2}$
$s^{-1}$.

\begin{figure*}[tbp]
\centering
\includegraphics[width=18.cm]{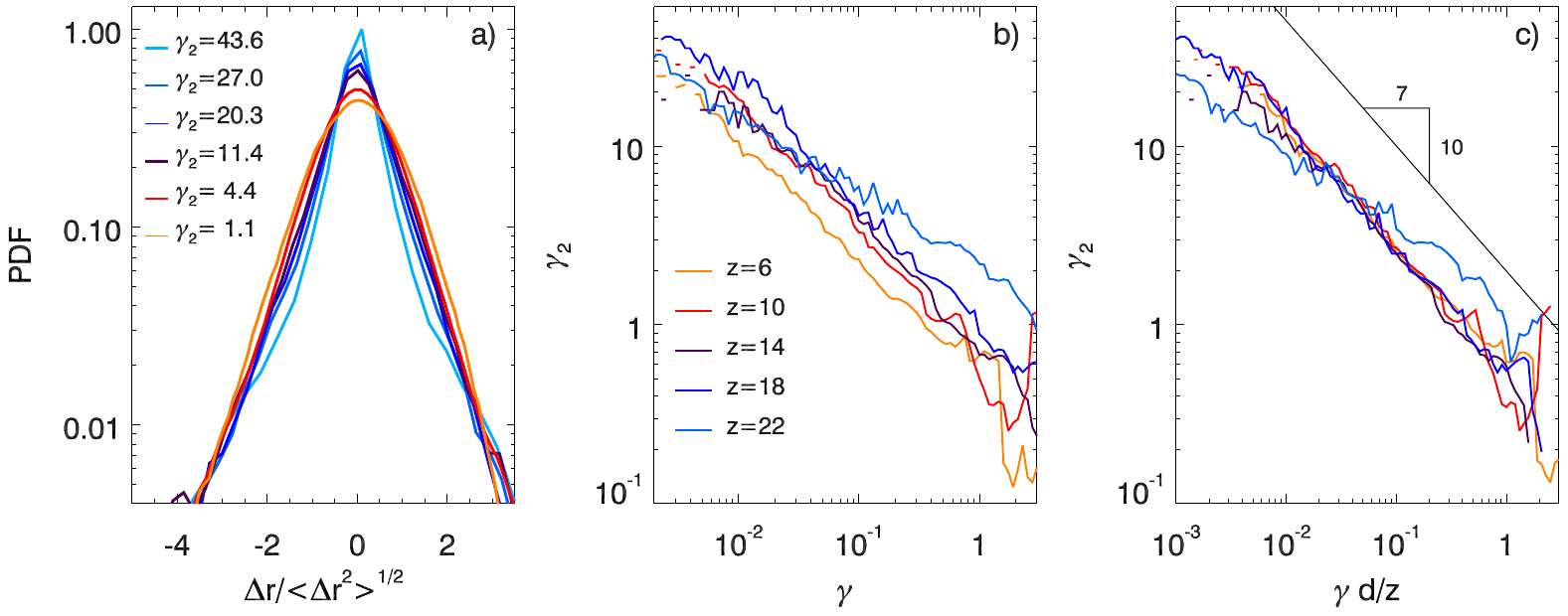}
\caption{ (a) PDF of radial displacements, for different values of
the reduced kurtosis $\gamma_2$, showing the range of displacement
distributions that we observe. (b) The radially averaged kurtosis
decays with the $\gamma$.(c) The radially averaged kurtosis
collapses when plotted as a function of the renormalized strain
$\gamma d/z$, and decays as a nontrivial powerlaw.
}
\end{figure*}

{\em Diffusivity ---} We have investigated the time evolution of
the non-affine part of the trajectory : $\Delta s_{\mu}(\Delta t)
= \Delta x_{\mu}(\Delta t) - \left< v_{\mu}\right > \Delta t $,
where $\mu$ denotes the component r, $\theta$ or $z$ of
cylindrical coordinates, and $\Delta x_{\mu}(\Delta
t)=x_{\mu}(t+\Delta t)-x_{\mu}(t)$. In the longitudinal ($\theta$)
direction and for small values of $r$ (see Fig.~1d), the mean flow
dominates, Taylor dispersion plays an important role, and reliable
data for the non-affine fluctuations are hard to obtain. We have
thus focused on the $z$ and $r$ components of the fluctuations for
large $r$, where the mean flow is slow (see Fig.~1d). While all
our data for the $z$-diffusion is consistent with the trends seen
for the $r$-diffusion, the spatial resolution in $z$ is less, so
we focus our presentation on the radial diffusion.

The normalized mean square displacements $\left<\Delta
s_{r}(\Delta t) ^2\right>/d^2$ are plotted as function of $\Delta
t$ in Fig. 2a, for two values of $z$, six values of $r$ and a
concomitant range of local strain rates which spans roughly two
decades. At late times, the observed behavior is diffusive, and
the mean squared displacement increases linearly with the time
lag. We thus define a diffusion coefficient $D$ as the limit for
large $\Delta t$ of the ratio $\left<\Delta s_{r}(\Delta t)
^2\right> /(\Delta t)$. For the cases shown in Fig.~2a, this
diffusion coefficient varies by two orders of magnitude.

In the simple local, quasi-static picture, the mean square
displacements would be determined by the local accumulated strain,
$\gamma:= \gammadot \Delta t$. Hence, one expects that these
curves would collapse when plotted as function of the strain.
Surprisingly, plotting $\left<\Delta s_{r}(\Delta t)
^2\right>/d^2$ as function of $\gammadot \Delta t$ does not
achieve to collapse the data into a single curve, but rather we
find two distinct curves, corresponding to data taken at the two
different $z$ positions (see Fig. 2b). The diffusivity thus
depends both on the strain rate and the vertical position in the
sample.

An alternative perspective on diffusivity starts from assuming a
Stokes-Einstein relation: $D=kT/(3\pi \eta d) $ where $D$, $\eta$
and $d$ are diffusion constant, effective viscosity and particle
diameter, and $kT$ refers to an effective temperature
\cite{privcom_ovarlez,MakseNature02}. For slow granular flows,
shear stresses $\tau$ are set by friction: $\tau = \mu P$, where
$\mu$ is the friction coefficient. As a consequence, the ratio of
shear stress and strain rate (the effective viscosity), scales as
$\eta=\mu P / \dot{\gamma}$, and substituting this into the
Stokes-Einstein relation, one obtains that
\begin{equation}\label{albertfrict}
D = (kT/3 \pi \mu d)( \gammadot/P)~,
\end{equation}
so that the diffusion would be inverse proportional to the local
granular pressure, which is set by the depth of the granular bed
$z$ (Eq.~\ref{albertfrict}).

As a non-dimensional characteristic of the depth of the granular
bed we take $z/d$ --- an integer that counts the number of grain
layers to the free surface. To obtain data collapse consistent
with the prediction $D \sim P^{-1}$, we can either multiply the
mean squared displacements $\left<\Delta s_{r}(\Delta t)
^2\right>$ with $z/d$, or multiply the local strain with $d/z$; as
we explain below, we believe the latter captures the physics best.
In Fig.~2c we show that all the mean squared displacements show
good data collapse when plotted as function of $\gammadot \Delta t
d/z$.

In Fig.~2d we combine our measured radial diffusivities for six
values of $r$, five values of $z$ and three global driving rates,
and find that in good approximation:
\begin{equation}
D \sim (\gammadot d^2)(d/z)~,
\end{equation}
with a proportionality constant of order one. If
the diffusion coefficient is plotted as a function of the strain
rate only, the diffusivities show a clear trend with $z$  (see
inset of Fig.~2d).

Whereas there is mounting evidence for the nonlocal nature of slow
granular flows \cite{NicholPRL10,PouliquenPRL10,GoyonNature08,
KamrinPRL12}; it appears that local plastic events affect the
viscosity in regions further away. One might have expected that
such nonlocal rheology would go hand in hand with a nonlocal
picture for diffusion, where the local diffusion constant is
governed by the strain rate in a finite region around the point
where the diffusion is probed. The good scaling collapse obtained
with purely local metrics suggests that nonlocal effects in
particle diffusion are weak or absent.

{\em Displacement Distributions ---} The rescaling of the
diffusion constant with inverse pressure leaves open one question:
is it fundamentally the strain scale or the displacement scale
that is affected by the pressure? To answer this question, we have
probed the probability distribution function (PDF) of the radial
particle displacements over a range of lag times $\Delta t$,
depths $z/d$ and local strain rates $\gammadot$.

While for late lag times the distributions become Gaussian, for
small $\Delta t$ we see appreciable deviations, with very wide
PDFs for the shortest lag times we can probe (See Fig.~3a).

We characterize the non-Gaussianity of the PDFs by the reduced
kurtosis $\gamma_2=\langle \Delta s_r^4 \rangle /{\langle \Delta
s_r^2 \rangle}^2 -3$, and determine the radial averages of the
kurtosis, which thus only depends on $z$ and $\gamma$. In Fig.~3b
and 3c we compare plots of the kurtosis $\gamma_2(z)$ for $z=6,
10, 14, 18$ and 22 mm, as function of strain $\gamma$ and rescaled
strain $\gamma d/z$. Clearly, the rescaled strain performs a
better rescaling of the kurtosis, even though the curve for
$z=22$, i.e. close to the bottom, appears to deviate from the
other curves. Together with our rescaling of $D$, this provides
strong evidence for a pressure dependent characteristic strain
$\gammadot \Delta t d/z$.

We finally note that on short time scales, the particle
displacement distributions appear to have fat tails. We note that
the kurtosis shows clean power law scaling with the rescaled
strain, with an exponent of approximately -0.7. Such power law
decay of the kurtosis may happen when the underlying stochastic
process is heavy-tailed~\cite{PottersEPL98}, and is a signature
that the distribution of individual jumps is of power law form.

{\em  Discussion ---}  Our results show that particle
displacements in sheared granular flow are diffusive on long time
scales, with a diffusivity proportional to the shear rate
\cite{BehringerPRE04,WeitzPRL10} and a surprisingly inverse
proportionality to depth.

A natural question concerns the microscopic origin of the
diffusion scale in slow granular flows. In a Langevin equation
type of approach, the order of magnitude of the diffusion
coefficient is $D \sim \frac{F_0}{\eta}$, where $F_0$ is the
stochastic force acting on the particles and $\eta=\frac{\mu
P}{\dot{\gamma}}$, the effective drag coefficient.

Considering the forces (gravitation, viscous, friction, inertial)
and the velocity scales (say $U<R_s\Omega$) in the present study,
we see that the viscous forces $F_v\sim3\pi \eta_f d U $ (where
$\eta_f$ is the fluid viscosity) and inertial forces $F_i\sim
\frac{\rho \pi d^3 }{6} \frac{U^2}{d}$ are negligible with respect
to gravitational $F_g=\frac{\pi d^3}{6} \Delta\rho g$ and
frictional forces $F_f \sim \mu P d^2$. In other words, the
Reynolds number is very small $Re=F_i/F_v \sim 10^{-4} $ and the
Stokes number is large $St=F_g/F_v\sim{10}$. We are in the "free
fall" regime described by Courrech du Pont et. al. in
\cite{CourrechPRL03} --- indeed, we expect our data to be relevant
for slow dry flows also.

This leaves frictional and gravitational forces as scales that set
the diffusion. Clearly, friction sets the drag, but it is not
obvious whether friction or gravity provides the dominant
stochastic force. If friction would dominate, both the driving and
damping would be set by it, and we would expect that the diffusion
coefficient would be independent of depth. In contrast, the
combination of a gravitational drive and frictional damping gives
the observed depth dependence of the diffusion as we show now.
Assuming that gravity sets the driving force for diffusion, we
find that:
\begin{equation}
D=\frac{\pi }{6}\dot{\gamma} d^2 \frac{\Delta\rho g d}{\mu p}\approx\dot{\gamma} d^2 \frac{d}{z}
\end{equation}
\noindent since $p\approx \Delta\rho g z$ and $\mu \approx 0.5$
\cite{DijksmanPRE10}. This expression is plotted as a solid line
in Fig 2d. The agreement is reasonable, supporting the idea of
gravity as a driving force for granular diffusion in slow flows.

\par Acknowledgements: The authors thank Jeroen Mesman for outstanding
technical support in the conception and construction of our 3D
scanner, Kinga L\"{o}rincz for her help in the image analysis, and
Bob Behringer, Daniel Bonn and Guillaume Ovarlez for fruitful
discussion. This work is part of the research programme of the
Foundation for Fundamental Research on Matter (FOM), which is part
of the Netherlands Organisation for Scientific Research (NWO).


\begin{thebibliography}{99}

\bibitem {WeitzPRL10} C. Eisenmann et. al., \textit{Phys. Rev. Lett.} \textbf{104} 035502 (2010)

\bibitem {SchallScience07} P. Schall, D. A. Weitz,  F. Spaepen, \textit{Science} \textbf{318} 1895  (2007)

\bibitem {NIPADurianPRL2010} K. N. Nordstrom et. al., \textit{Phys. Rev. Lett.} \textbf{105} 175701 (2010)

\bibitem {GoyonNature08} J. Goyon et. al. \textit{Nature}  \textbf{454}, 84-87 (2008)

\bibitem {KatgertPRL08} G. Katgert, M. Mobius, M. van Hecke \textit{Phys Rev Lett} \textbf{101}, 058301 (2008)

\bibitem {TighePRL10} B. P. Tighe et. al. \textit{Phys. Rev. Lett.} \textbf{105}, 088303 (2010)

\bibitem {ForterreAnnRevFlu08} Y. Forterre and O. Pouliquen \textit{Annu. Rev. Fluid Mech.} \textbf{40} 1 (2008)

\bibitem{GDRMidi} $GDR$ Midi \textit{E. Phys. J. E} \textbf{14} 367-371 (2004)

\bibitem {BehringerPRE04} B. Utter and B. Behringer  \textit{Phys. Rev E} \textbf{69} 031308 (2004)

\bibitem {LosertPRL00} W. Losert et. al. \textit{Phys. Rev. Lett} \textbf{85} 1428 (2000)

\bibitem {PouliquenPRL04} O. Pouliquen \textit{Phys. Rev. Lett.} \textbf{93} 248001 (2004)

\bibitem {FenisteinPRL06} D. Fenistein, J.-W. van de Meent and M. van Hecke \textit{Phys. Rev. Lett} \textbf{96} 118001 (2006)

\bibitem {DijksmanPRE10} J.A. Dijksman et. al.  \textit{Phys. Rev. E } \textbf{82}, 060301(R) (2010)

\bibitem {BoyerPRL11} F. Boyer, E. Guazzelli and O. Pouliquen \textit{Phys. Rev. Lett} \textbf{107}, 188301 (2011)

\bibitem {DijksmanPRL11} J.A. Dijksman et. al. \textit{Phys. Rev. Lett} \textbf{107}, 108303 (2011)

\bibitem {PouliquenPRL10} K. A. Reddy, Y. Forterre and O. Pouliquen \textit{Phys. Rev. Lett} \textbf{106}, 108301 (2011)

\bibitem {NicholPRL10} K. Nichol et. al. \textit{Phys. Rev. Lett} \textbf{104}, 078302
(2010); K. Nichol and M van Hecke, Phys. Rev. E. {\bf85} 061309
(2012).

\bibitem {KamrinPRL12} K. Kamrin and G. Koval \textit{Phys. Rev. Lett} \textbf{108}, 178301 (2012)

\bibitem{clement08} A. Amon, V B Nguyen,  A. Bruand, J. Crassous and E
Cl\'ement, Phys. Rev. Lett. {\bf 108} 135502 (2012)

\bibitem {MuethNature2000} D. M. Mueth et. al. \textit{Nature} \textbf{406}, 385 (2000)

\bibitem {SchallARFM10} P. Schall and M. van Hecke \textit{Ann. Rev. Fluid Mech.} \textbf{42}, 67--88 (2010)

\bibitem{CourrechPRL03} S. Courrech du Pont et. al. \textit{Phys. Rev. Lett.} \textbf{90} 044301 (2003)

\bibitem {DijksmanSM10} J.A. Dijksman and M. van Hecke \textit{Soft Matter} \textbf{6} 2901 (2010)

\bibitem {TsaiPRL03} J.C. Tsai, G. Voth and J. Gollub \textit{Phys. Rev. Lett.} \textbf{91} 064301 (2003)

\bibitem {LosertPRL08} S. Slotterback et. al. \textit{Phys. Rev. Lett} \textbf{101}, 258001 (2008)


\bibitem {KingaSoftMatt10} K. Lorincz and  P. Schall \textit{Soft Matter} \textbf{6}, 3044-3049 (2010)


\bibitem {DijksmanRSI12} J. A. Dijksman et. al.  \textit{Rev. Sci. Instr} \textbf{83} 011301 (2012)


\bibitem {WeeksTrack96} J. C. Crocker and D. G. Grier \textit{J. Coll. Int. Sci.} \textbf{179} 298 (1996)


\bibitem {privcom_ovarlez} G. Ovarlez, \textit{private communication}

\bibitem {MakseNature02} H. A. Makse and J. Kurchan \textit{Nature} \textbf{415}, 614 (2002)


\bibitem {MobiusEPL10}  M. Mobius, G. Katgert, M. van Hecke \textit{EPL} \textbf{90} 54002 (2010)


\bibitem {PottersEPL98} M. Potters, R. Cont and J.-P. Bouchaud \textit{EPL} \textbf{41} 239-244 (1998)


\end{thebibliography}
\end{document}